\documentclass[aps,twocolumn,preprintnumbers,amsmath,amssymb,floatfix]{revtex4}
\usepackage{wrapfig}
\usepackage{times}
\usepackage{epsfig}
\usepackage{graphicx}
\usepackage{color}
\usepackage{bm}
\usepackage{nomencl} 
\makenomenclature
\usepackage{amsmath}
\usepackage{amssymb}
\usepackage{setspace}
\usepackage{pdflscape}

\newcommand \beq {\begin{equation}}
\newcommand \eeq {\end{equation}}
\newcommand \beqa {\begin{eqnarray}}
\newcommand \eeqa {\end{eqnarray}}
\newcommand{\sbeq}{\begin{subequations}}
\newcommand{\seeq}{\end{subequations}}

\newcommand \nline {\nonumber \\}

\newcommand \freem {{\cal F}}

\newcommand \pxpy[2] {\frac{\partial #1}{\partial #2}}
\newcommand \dxdy[2] {\frac{d #1}{d #2}}

\begin{document}

\title{Microscopic Treatment of Solute Trapping and Drag}

\author{Harith Humadi$^{1,2}$}
\author{J. J. Hoyt$^2$}
\author{Nikolas Provatas$^{1}$}

\affiliation{$^1$Department of Physics, Centre for the Physics of Materials, McGill University, Montreal, QC, Canada\\$^2$ Department of Materials Science and Engineering, McMaster University, Hamilton, Ontario }

\date{\today}

\begin{abstract}
 The long wavelength limit of a recent microscopic phase field crystal (PFC) theory of a binary alloy mixture is used to derive an analytical approximation for the segregation coefficient as a function of the interface velocity, and relate it to the two-point correlation function of the liquid and the thermodynamic properties of solid and liquid phases. Our results offer the first analytic derivation  of solute segregation and solute drag derived from a microscopic model, and analytically support recent molecular dynamics and fully numerical PFC simulations. Our analytical result also provides an independent framework, motivated from classical density functional theory, from which to elucidate the fundamental nature of solute drag, which is still highly contested in the literature.  
\end{abstract}

\maketitle

There are many theories explaining the morphologies and the underlying physics for near-equilibrium systems that evolve towards their equilibrium state~\cite{langer99}. By contrast, theories of physical phenomena associated with far-from-equilibrium systems remain much less developed. Rapid solidification from highly undercooled melts serves as a paradigm of such phenomena. In processes like laser-induced surface melting, spray forming, and welding among other technologies, highly super-saturated meta-stable solid solutions can form. In many cases, the non-equilibrium nature of such process can be exploited to control the degree of super-saturation of the solid. 

At rapid-solidification rates, solute concentration at the solid-liquid interface (SLI) can deviate substantially from the values predicted by the equilibrium phase diagram, a phenomenon known as {\it solute trapping}~\cite{kittl00, aziz88, Galenko97b, Ahmad98, jackson04, sob95, sob97}, 
In addition to solute trapping, the growth of a crystal with a composition differing from that of its melt requires solute diffusion to move across the SLI. The free-energy dissipation associated by interface diffusion leads to the phenomenon of {\it solute drag}, an effect which
 can strongly hinder the transformation rate.  Solute drag arises due to a competition between interface diffusion rate and a chemical
 potential difference across the interface. When the velocity of the SLI is low, local equilibrium is assumed, the chemical potential difference between the SLI essentially vanishes, and solute drag is negligible. As the interface speed increases,  solute diffusion limits the rate of partitioning across the interface (solute trapping), leading to an increasing chemical potential jump with velocity and, hence, an increasing solute drag. At large SLI speeds, solute partitioning eventually stops, as does diffusion of solute through the interface, and thus solute drag vanishes. 

A phenomenology of solute drag was proposed in the seminal work by Cahn \cite{Cahn62} for the case of a grain boundary separating two solid phases. Although the Cahn model quantitatively predicts various aspects of the drag effect,
it was assumed that the chemical potential is equal on both sides of the transformation front, an assumption that does not hold for a rapidly solidifying front. Later, Hillert and Sundman~\cite{Hil76} incorporated a chemical potential jump into their phenomenology, and proposed that the maximum amount of free energy associated with drag is dissipated.  A model for solute drag for solidification was first proposed by Hillert~\cite{Hillert00}, which considered the structure of the interface and its effect on drag. Solute drag experiments are difficult to perform. Some  show a significant change in solute concentration at the SLI interface at rapid solidification rates \cite{Smith94}, while  some ~\cite{kittl95} even find no evidence of solute drag. Subsequent models  proposed a partial solute drag hypothesis~\cite{Aziz94,Ahmad98,yangprl, Agren89}.  More recently, atomistic simulations of Yang et al~\cite{yangprl} and Humadi et al~\cite{Humadi13} proved that the solid-liquid interface stops partitioning solute at a finite velocity, consistent with predictions of Sobolev et al~\cite{sob95, sob97} and in contrast to earlier predictions of Aziz et al~\cite{aziz88,Aziz94} with some evidence of partial solute drag. 

Traditional phase field models of solidification consider bulk mass and heat transport coupled to moving interfaces through effective equilibrium boundary conditions \cite{Cag89,Cag91,Karma98,Provatas98,Kim99b,Karma01,Boettinger02,Echebarria04} that map onto traditional sharp interface models. While such an approximation is appropriate at low solidification rates, it is inappropriate at rapid cooling rates where, as described above, non-equilibrium solute partitioning and drag become dominant. Based on the pioneering works of Cahn and Hillert \cite{Cahn62,Hil76,Hillert00}, modified sharp interface models were developed for rapid solidification~\cite{Azizbot93,Aziz94}. However, these models are largely phenomenological and are based on physically motivated, but often ad-hoc, parameters that cannot link the solidification kinetics to any microscopic quantity of the liquid and solid.  More recent phase field modelling of rapid solidification has confirmed much of the phenomenology of these sharp interface models~\cite{kittl00, aziz88, Galenko97b, Ahmad98, jackson04, sob95, sob97}.  Still, no fundamental link between the meso-scale solidification process and the microscopic parameters of the materials can be made since solute trapping and drag fundamentally emerge at the atomic scale, where traditional phase field models, by their very nature, lack any qualitative and quantitative detail  \cite{Echebarria04,ProvatasElder10}. {\it At present, no microscopic treatment of the trapping and solute drag coefficients entering rapid solidification models exists}.

Recently, an emerging atomistic continuum modelling formalism coined the {\it phase field crystal} (PFC) method has been developed that presents an alternate atomistic framework with which phenomena such as solute trapping can be studied. In contrast to the traditional phase-field approach, PFC models are formulated in terms of order parameters that are periodic at the atomic scale, but whose dynamics evolve over diffusive time scales relevant to rapid solidification processes. A phase field crystal model of binary alloy solidification was first derived in Ref.~\cite{Elder07} as a simplification of a truncated density functional theory (DFT) expanded around the liquid state at coexistence. As such, the model inherits crucial microscopic liquid state parameters originating from the two-point correlation function of the solidifying liquid. 
The approach has been shown to self-consistently incorporate the physics of nucleation, multiple crystal orientations, grain boundary energy, elasto-plascitiy and topological defects and their dynamics~\cite{Elder02,Elder04,Berry06,Stefanovic06,Mellenthin08,Stefanovic09,Granasy10,granasyprl}.  A significant advance in PFC modelling is its use with multi-scale and renormalization methods to project out meso-scale phase field models with complex order parameters \cite{Athreya06,Athreya07,Huang10}, the coefficients of which maintain their connection to the microscopic liquid and solid state properties inherent in the generating PFC theory. {\it In this letter}, we use a PFC-dervied amplitude model of solidification to elucidate, for the first time, an analytical derivation of the non-equilibrium solute partition coefficient and the solute drag coefficient that enters models of solute drag. 

Multiple scale analysis applied to the PFC alloy model in \cite{Elder07} yields the following moving front equations for the impurity concentration ($\psi$) and the amplitude of the reduced atomic number density $(\phi$) \cite{Humadi_thesis},
\beqa
\beta V^2 \dxdy{^2\phi}{z^2}-V\dxdy{\phi}{z}\!\! &=&\!\! W^2(\hat{n}) \dxdy{^2\phi}{z^2}-\pxpy{f}{\phi} \label{amp_eqs} \\ 
\gamma V^2 \dxdy{^2\psi}{z^2}-V\dxdy{\psi}{z}\!\!&=&\!\! \dxdy{}{z}\bigg(M\dxdy{}{z}\bigg\{(\omega+6B_2^{\ell} \phi^2)\psi+u\psi^3 \bigg \} \!\bigg) \nonumber
\eeqa
Their derivation assumes that the atomic number density $n\equiv (\rho-\bar{\rho})/\bar{\rho}$ is represented by $n=n_0+\sum_j A_j e^{i G_j \cdot \vec{x}}$, where $n_o$ is the reduced average alloy density, and $\bar{\rho}$ is the reference liquid density at coexistence. It is assumed that $n_0=0$ here for simplicity. The $\vec{G}_j$ is the $j^{th}$ reciprocal lattice vector of a general multi-mode expansion of the density, and $A_j$ is the complex density amplitude corresponding to the $j^{th}$ density wave. We consider here a 2D triangular crystal structure but
the qualitative physics of our results are not expected to change for other crystal structures. For solidification, it is suitable to set all the $A_j$ to be real, i.e. $A_j=\phi$.  The equations are written in a co-moving 1D reference frame moving at velocity $V$, which is accurate for rapid solidification. The second order derivatives allow for a two-time scale relaxation of the density and concentration fields. They can be motivated by considering mass and momentum conservation of two-species densities $\rho_A$ and $\rho_B$ \cite{Stefanovic06,Majaniemi07,Humadi13}. The coefficients $\gamma$ and $\beta$ are microscopic relaxation parameters for the solute and density, respectively, while $M$ is the mobility of impurity atoms. The variable $W(\hat{n})=B_0^x \sum_j \hat{n} \cdot \vec{G}_j $, where $\hat{n}$ is the local interface normal vector and $B_o^x$ is the lowest order coefficient of the solid compressibility. The liquid compressibility is denoted by $B^l$ and expanded as $B^l=B_0^l+B_2^l \psi^2$ \cite{Elder07}.  The bulk free energy is denoted by $f(\phi,\psi)$ and 
$\partial f/\partial \phi=6[\Delta B_o +B_2^l \psi^2]\phi -12t\phi^2 +90\nu \phi^3$, where $\Delta B_0=B_0^l-B_0^x$. The variables 
$t,\nu,\omega,u$ are the respective coefficients of the bare $\phi^3$, $\phi^4$, $\psi^2$ and $\psi^4$ terms of a landau expansion of the bulk free energy. Bulk compressibility of the liquid  $B^l=1-\bar{\rho} \hat{C}_0$, and $B_0^x=\bar{\rho}\hat(C)_2^2/(4\hat{C}_4)$, where $\hat{C}_2, \hat{C}_2, \hat{C}_4$ are coefficients of a fourth order expansion of the two-point correlation function of the liquid state, given by $C({\bf k})=\hat{C}_0+\hat{C}_2 k^2+\hat{C}_4 k^4$ \cite{Elder07}.  In what follows, we rescale $\bar{\phi}= \phi/\phi_s$ and $\bar{\psi} = \psi-\psi_s$, where $\phi_s$ and $\psi_s$ are the bulk order parameter and concentration of the solid phase, respectively. All results presented here are for $\{\nu,t,u,\omega,B_2^l,B_o^x,W(n),\phi_s,M\}=\{1,0.6,4,0.008,-1.8,1,2,0.06,1\}$.

For the parameters above, the equilibrium partition coefficient of the PFC model of Ref.\cite{Elder07} is $K_e=0.97$. The 
solidus-to-liquidus jump for this model is $\epsilon =(\psi_s+1)/K_e -(\psi_s + 1 ) \ll1$,  which forms an ideal small parameter 
to expand  $\psi$. 
Notice that in the PFC model, with a quite high value of $K_e$, the paramter $\epsilon$ is indeed very small.  Nevertheless, the jump
in concentration between liquid and solid is small in most alloy system and thus we anticpate the results derived below to be 
applicable in general. 
Integrating  the $\psi$ equation in Eq.~\ref{amp_eqs}  from $-\infty$ to $z$ and substituting 
$\psi\approx \psi_s +\epsilon \psi_1 +\epsilon^2 \psi_2 + \ldots$ into the result gives the following 
$\mathcal{O}(\epsilon)$ equation for $\psi_1$,
\begin{equation}
\gamma V^2 \dxdy{\psi_1}{z} \!-\! V\psi_1\!=\! M \dxdy{}{z} \!\left( \! \bigg[\omega \!+\! 6 B_2^{\ell}\phi_s^2 \phi^2 + 3u\psi_s^2\bigg] \psi_1\right)
\label{order_e_psi}
\end{equation}
Equation~\ref{order_e_psi} will be used to approximate the concentration profile in the liquid. Higher order terms are needed to approximate concentration in the solid, but that will not be necessary here and will be omitted in what follows.

In Eq. 1 the parameter $W$ is a measure of the SLI width and therefore we approximate  the order parameter $\phi  \approx \phi_o(z) \equiv [1\!-\! \tanh(z/W)]/2$, and define $z=0$, where $\phi_o(0)=1/2$, as the interface between solid and liquid ordering analogous to molecular dynamics studies \cite{yangprl}. $\phi_o(z)$ is is the exact lowest order solution of the PFC model for a pure material \cite{private_galenko}. We have found that it is also a reasonable approximation for the density amplitude of the PFC alloy model.    Substituting the above expression for $\phi$ into Eq.~(\ref{order_e_psi})  gives,
\beqa
- \frac{V}{M} \psi_1 = \dxdy{}{z} \bigg[ b + \delta \bigg[1 - \tanh \bigg( \frac{z}{W} \bigg) \bigg]^2 \bigg] \psi_1
\eeqa
where $ b\equiv \omega + 3u\psi_s^2 - \gamma V^2/M$ and $\delta \equiv 6 B_2^{\ell}\phi_s^2$. This equation can be solved analytically with an integrating factor that must be solved numerically. In favour of obtaining a tractable analytic expression to work with, we exploit the fact that 
$\delta/b \sim 10^{-2}$ and $|\tanh| <1$ and seek an analytical solution to lowest order in $\delta/b$. This gives,  
\beq
\psi_1 \approx  \large{ e^{ - \frac{V}{Mb} \left\{ z +  \frac{\delta W}{b} \Phi_o(z) \right\} \, } }
 \left\{1+ \mathcal{O}\left(\delta/b\right)\! + \!\cdots \! \right\}
\label{lquidconcprof} 
 \eeq
where $\Phi_o(z) \equiv   \tanh\left(\frac{z}{W}\right) \!-\! 2 \ln \left(1 + \tanh\left(\frac{z}{W}\right)\, \right)$.
In obtaining Eq.~(\ref{lquidconcprof}), the integration constant was found by applying the boundary condition $\psi(z=W/2) = \psi_{\ell}^{e}\equiv\epsilon+\psi_s$ at $V=0$, where $\psi_{\ell}^e$ is the equilibrium liquid concentration, and $z=W/2$ defines the point where the concentration profile reaches a maximum, 
consistent with molecular dynamics \cite{yangprl} and previous PFC alloy simulations \cite{Humadi13}. We also take the far field concentration in the liquid to be the same as the solid concentration $\psi_s$. For simplicity, we analyze only the exponential part of Eq.(\ref{lquidconcprof}). We found that including the higher order terms gives essentially the same results. 

\begin{figure}[htbp]
\centering
	\resizebox{3.0in}{!}{\includegraphics{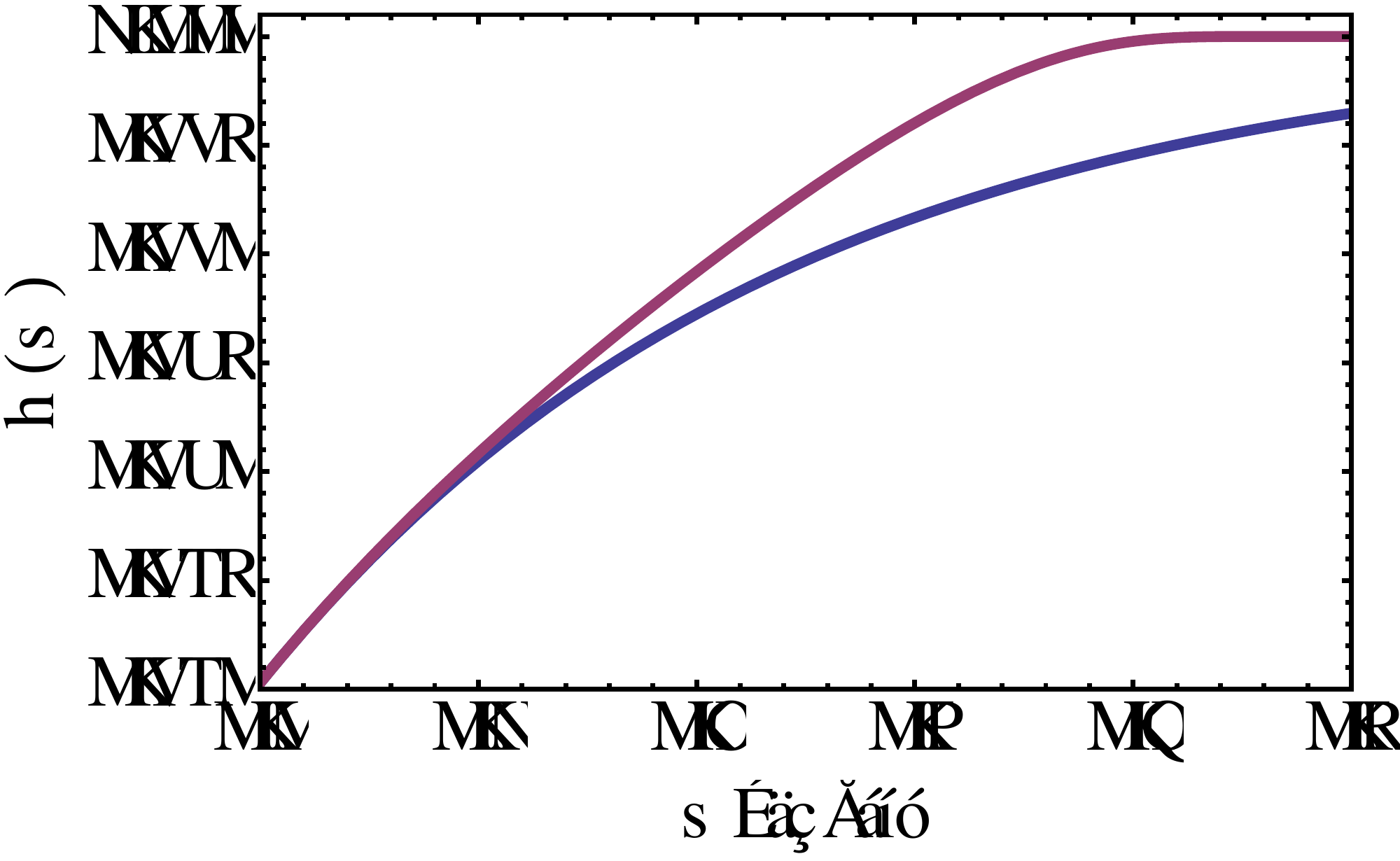}}
	\caption[Analytical Segregation Coefficient]{Segregation coefficient as a function of the interface velocity. The blue line represents the diffusive case where the $K(V)$ tends to unity asymptotically as $V \rightarrow \infty$. The purple line illustrates the case where the $K(V)$ reaches unity at finite velocity, here $V\approx 0.4$.}
\label{kvelamp}
\end{figure}

The segregation coefficient $K(V)$ is defined to be the ratio of the interface solid concentration to that of the maximum liquid concentration, which occurs when $\phi=1/2$ at $z=W/2$. In the PFC model, the concentration is expanded around $c=0.5$, which yields negative concentrations on the left side of the phase diagram. As a result, the solute  partition coefficient for the PFC alloy model is defined as
\beq
K(V)=\frac{ \psi_s+1}{ \left(\psi_s+\epsilon \psi_1(W/2)\right) + 1 }
\eeq
Figure~(\ref{kvelamp}) plots $K(V)$ for two cases, the first case (purple curve online) for $\gamma \ne 0$ and the second  case (blue curve online) for $\gamma=0$. For the first case $K(V)=1$ at a {\it finite} $V$, while in the second case, $K(V) \rightarrow 1$ only asymptotically as the solid-liquid interface velocity $V \rightarrow \infty$. 

There are two competing theories for explaining $K(V)$ in the literature. The first, by Aziz~\cite{aziz88} assumes purely diffusive solute transport and flux balance across the interface to predict the segregation coefficient. Aziz predicts that $K(V)$ approaches complete trapping ($K(V)=1$) asymptotically, and never reaches unity at finite $V$. More recently, Sobolev~\cite{sob95,sob97} proposed a phenomenology that considered inertial dynamics of solute atoms in the liquid. This lead to the emergence of an effective diffusion coefficient, which makes it possible for $K(V)$ to reach unity at a finite velocity. 

In our microscopic PFC formalism, the constant $b$ in Eq.~\ref{lquidconcprof} emerges as an effective diffusion coefficient. The value of $b$  decreases to zero as the interface velocity increases. As a result, the liquid concentration tends to the solidus concentration. However, this is only true for non-zero inertial solute relaxation time ($\gamma \neq 0$). Otherwise, $b$ always remains non-zero, and does not change the classical diffusive nature of the concentration profile. This allows for a concentration jump to develop across the two sides of the interface, even for arbitrarily large interface velocities($V$). 

The above analytical PFC result is consistent with the previous numerical simulations of the alloy PFC model~\cite{Humadi13} as well as with recent molecular dynamics simulations \cite{yangprl}. We note that a higher order perturbation analysis of the coupled Eqs.~(\ref{amp_eqs}) would be required to compare the results quantitatively with the full numerical simulations. However, the physics does not change. {\it Our result is the first  solute trapping theory to offer a prediction of the complete solute trapping velocity in terms of  microscopic parameters}. Namely, Eq.~(\ref{lquidconcprof}) predicts that complete trapping occurs when $V^*=\{M\left(\omega +3u \psi_s^2\right)/\gamma\}^{1/2}$. The approximate form of $\psi_s$ was derived in Ref.~(\cite{Elder07}), given by: $\psi_s \! \approx \! \pm \psi_{sl} \left( 1+ G \{ 1- \sqrt{b_{\rm liq}/b_{\rm sol}} \} \right)$, where the variables in this expression are given by $\psi_{sl}\!=\! \sqrt{\left(\Delta B_o^{sl}- \Delta B_0\right)/B_2^l}$, $G=-8t^2/\{135v\left(4\Delta B_0 - 3\Delta B_o^{sl}\right) \}$, $\Delta B_o^{sl}=8t^2/135v$, while  $b_{\rm liq}=\left(\omega + 3u\psi_{sl}\right)/2$ and $b_{\rm sol}= b_{\rm liq}+2B_2^l \left(4 \Delta B_0 -3 \Delta B_o^{sl}\right)/5v$. Thus, we have shown that the complete trapping velocity is inversely proportional to the square root of the inertial relaxtion tme and proportion to $\psi_s$, which is determined by the properties of the two-point correlation function of the liquid $C(|{\bf k}|)$, through $B_l, B_o^x$, and the bulk solid free energy density, through ($t,v,\omega,u$).

Solute drag in the context of the PFC formalism can also be elucidated using Equation~(\ref{lquidconcprof}). The theoretical formalism of solute drag is briefly summarized here. The free energy density available for solidification of a binary alloy (denoted here as $\Delta G_s$) is partially dissipated due to solute atoms diffusively redistributing parallel to the solidifying front before  attaching to the solid phase. This dissipation is referred to as solute drag, and reduces the total effective free energy available for solidification (denoted $\Delta G_c$) to  
\beq 
\Delta G_c = \Delta G_s - f \Delta G_d
 \label{xtalfreeng}
\eeq
where the maximum drag was shown by Ahmad et al to be $\Delta G_d = (\psi_{\ell} - \psi_s)( \mu_{\ell} -  \mu_s)$~\cite{Ahmad98}, while $\Delta G_s= \freem_s(\psi_s,T) -  \{ \freem_{\ell}(\psi_{\ell},T) + (\psi_s - \psi_{\ell}) ( \mu_{\ell}) \}$, derived by  Cahn~\cite{Cahn71}, where $\mathcal{F}$ denotes bulk free energy density and $\mu_{\ell}$ and $\mu_s$ are the inter-diffusional chemical potentials of the liquid and solid phase and evaluated at $\psi_l$ and $\psi_s$, which are, respectively, the liquid and solid concentrations on the liquid and solid side of the interface.  We can equivalently express $\Delta G_s = \psi_s \Delta \mu_B + (1 - \psi_s) \Delta \mu_A$, where $\Delta \mu_A$ ($\Delta \mu_B$) are the solvent $A$ (solute $B$) chemical potential differences between the solid and liquid phases. The important constant $f$ has limits $0<f<1$, with $f=0$ implying zero drag $f=1$ maximum drag. We determine $f$ in the PFC formalism below.

The above expressions for $\Delta G_s$ and $\Delta G_d$ were applied by Ahmad and co-workers~\cite{Ahmad98} to a phenomenological phase-field model. Since the PFC amplitude equations~(\ref{amp_eqs}) are also a phase-field theory, derived by coarse graining a microscopic PFC theory, we similarly apply the above expressions to the free energy of the PFC amplitude model. This is derived from $f(\phi,\psi)$, which in the bulk gives   
\begin{align}
\freem_s&=\frac{45 \nu \phi_s^4}{2}-4t \phi_s^3+3 \left(B^{\ell} \!\!-\!\! B_o^x\right) \phi_s^2  +\frac{u \psi_s^4}{4}+\frac{\omega \psi_s^2}{2} \nline
\freem_{\ell}&=\frac{u \psi_l^4}{4}+\frac{\omega \psi_l^2}{2}
\label{SLampenergies}
\end{align}
for the free energy density in the solid ($\freem_s$) and liquid ($\freem_{\ell}$).

\begin{figure}[htbp]
\centering
	\resizebox{3in}{!}{\includegraphics{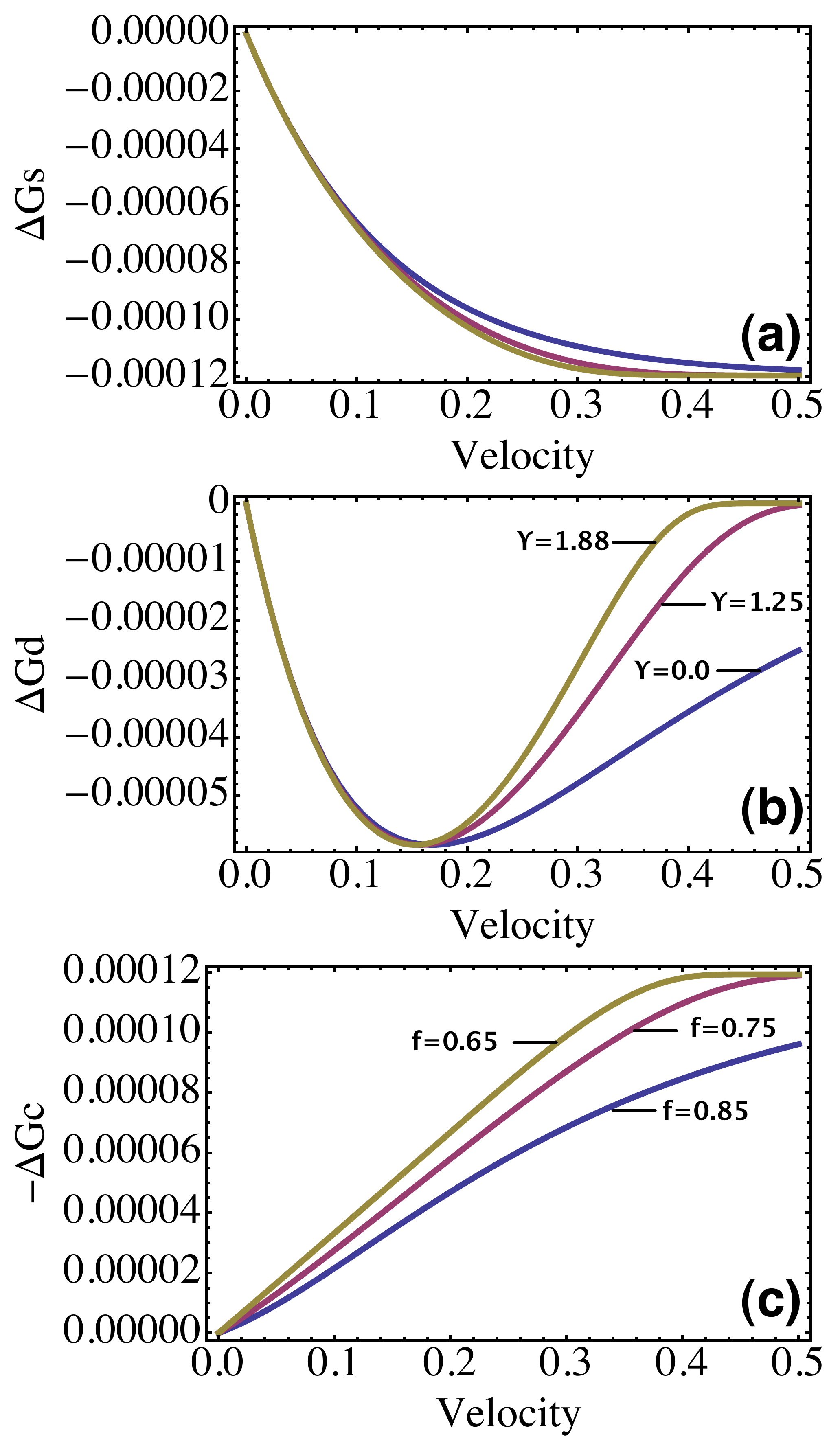}}
	\caption[Effect of $\gamma$ on Solute Drag]{The driving forces for crystallization for $\gamma=0,1.25, 1.88$, where the colour (online) corresponding to each $\gamma$ is shown in panel (b)). (a) The total driving force available for transformation. (b) The maximal solute drag. (c) The total available crystallization free energy vs. interface velocity. Each curve ($\gamma$ value)  shows the best fit $f$ that makes 
	$\Delta G_c \propto V$ at low velocities. }
\label{fig:DG_gamma}
\end{figure}

At low thermodynamic driving forces, molecular dynamics simulations and experiments suggest that $V \propto - \Delta G_c$ ~\cite{turnbull62}, a relation that becomes less accurate near complete trapping velocities. The solute drag coefficient $f$ in Eq.~(\ref{xtalfreeng})   can thus be determined by tuning $f$ until a linear relationship between $V$ and $\Delta G_c$ emerges. We do so here numerically. 
To proceed, the solid concentration $\psi_s$ and order parameter $\phi_s$ are assumed constant in the solid during steady-state front propagation, while the liquid concentration $\psi=\psi_s+\epsilon \psi_1$ is determined by Eq.~\ref{lquidconcprof}. These quantities are substituted into $\freem_{\ell}$ and $\freem_s$ to compute $\Delta G_s$, $\Delta G_c$ and $\Delta G_d$.   Fig~\ref{fig:DG_gamma}a shows three different cases of $\Delta G_s$ versus $V$. The blue line represents the diffusive case where no complete trapping occurs ($\gamma=0$). The purple and the yellow lines show $\Delta G_s$ for $\gamma=1.24$ and $\gamma=1.88$, respectively. 
Fig~\ref{fig:DG_gamma}b plots $\Delta G_d$ for the same $\gamma$ values as Fig~\ref{fig:DG_gamma}a. It is noteworthy that the maximum amount of solute drag (minimum of $\Delta G_d$) does not change as the degree of trapping ($\gamma$) changes. However, the curvature of $\Delta G_d$ at large $V$ is quite sensitive to $\gamma$.  This occurs because as $\gamma$ increases, complete trapping occurs at lower velocity ($V^*$). This causes the concentration difference across the interface to decrease rapidly for $V>V^*$, thus leading to a decrease in $\Delta G_d$, which depends on $\psi_{\ell}-\psi_s$. Fig~\ref{fig:DG_gamma}c shows that for each $\gamma$,  the partial  solute drag fraction $f$ exists for which $\Delta G_c \propto V$ at low SL velocities. This confirms previous solute drag phenomenologies, and is consistent with recent molecular dynamics results~\cite{yangprl}. Our results illustrate that as the solute relaxation coefficient $\gamma$ changes $V^*$, and the degree of solute trapping, it also affects the driving force for complete crystallization through $\Delta G_d$ and the solute drag coefficient $f$. 

\begin{figure}[htbp]
\centering
	\resizebox{3in}{!}{\includegraphics{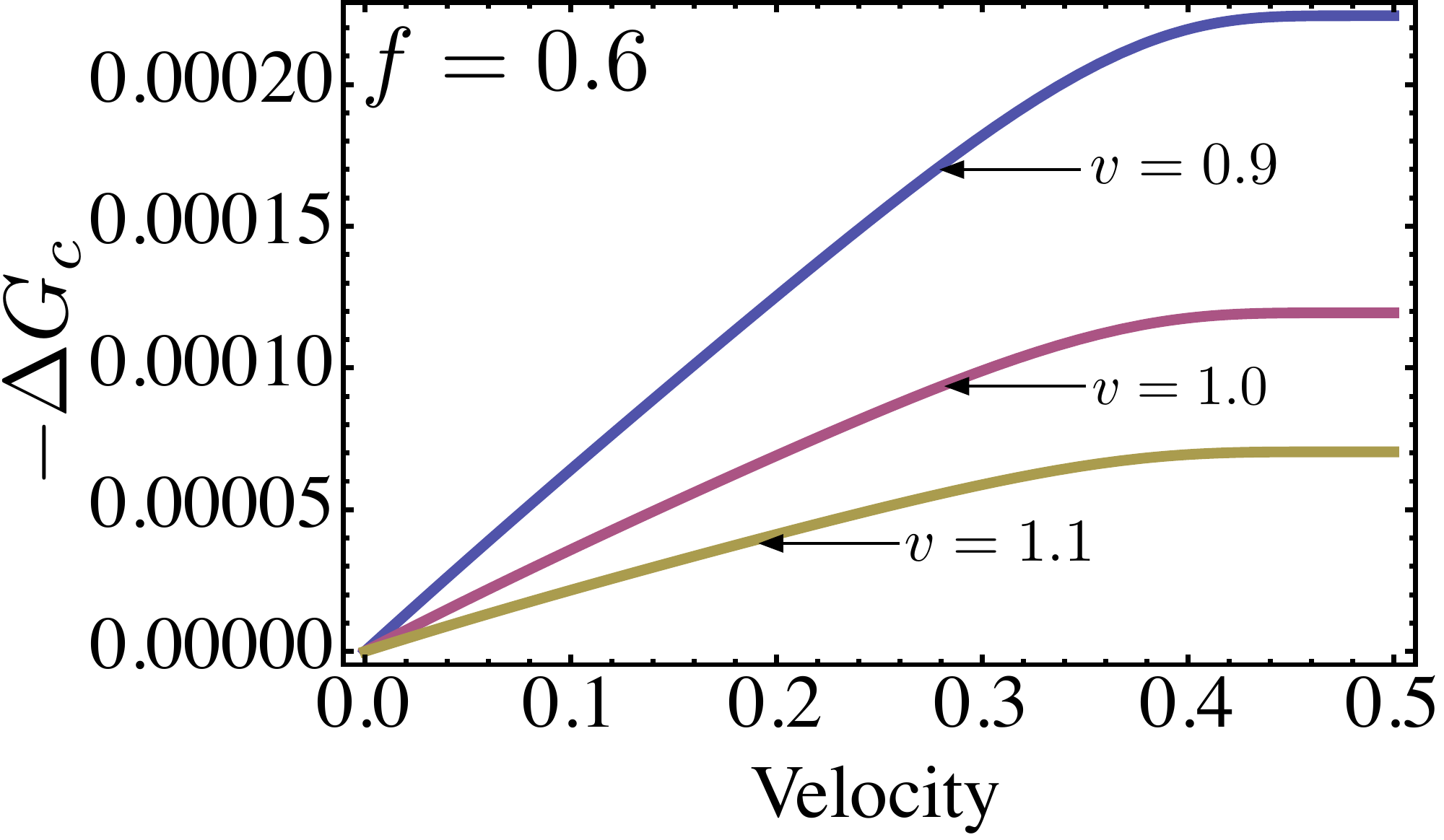}}
	\caption[Thermodynamic Effects on Solute Drag]{The driving force for crystallization, the three different lines show the different $K_e$ by changing the value of $\nu$ ($\gamma=1.88$ is fixed).}
	\label{fig:DG_v}
\end{figure}

Other materials parameters of our phase field crystal theory were also examined for their effect on solute drag.  An important one is the equilibrium solute partition coefficient $K_e$, which is controlled by $\nu$, the coefficient of the $\phi^3$ term in the bulk PFC free energy functional. Increasing $\nu$ leads to increasing $K_e$. Materials with larger $K_e$ exhibit lower complete trapping velocities ($V^*$) because less driving force is required to reach complete trapping for a decreasing concentration jump $\psi_{\ell}-\psi_s$. Thus, solute drag $\Delta G_d$ 
also decreases as $K_e$ increases. Interestingly, while $K_e$ changes the maximum available solute drag ($\Delta G_d$), we found that it does not change the partial solute drag coefficient $f$. Fig~\ref{fig:DG_v} illustrates $-\Delta G_c$ Vs. $V$ for three values of $\nu$ (or, equivalently, $K_e$), at a fixed $\gamma$ (other parameters are as indicated at the beginning of this paper). For all curves in  Fig~\ref{fig:DG_v}, the value of $f$ is the same. This illustrates that in all cases, the driving force for crystallization ($\Delta G_c$) increases as solute drag decreases because of the decreasing of $\Delta G_d$, {\it not} because $f$ is changing. This important prediction implies that solute drag is strictly a kinetic process (i.e. $\Delta G_d$ depends on $V$, through $K_e$) and that $f$ is a thermodynamic quantity that has no effect on the maximum solute drag. 

In summary, an amplitude model derived from a microscopic phase field crystal model has been derived to study the phenomena of solute trapping and solute drag, two important materials processes that remain poorly understood.  We have derived a first order analytic expression for the concentration profile in the liquid as a function of interface velocity and position, and used it to derive the solute trapping coefficient $K(V)$. Our model predicts that when inertial dynamics are included in solute transport, complete trapping occurs at a finite velocity, consistent with the phenomenology of Sobolev~\cite{sob95, sob97} and recent MD simulations. A key result is the derivation of an expression for the complete trapping velocity as a function of the bulk compressibility of the solid and liquid and the bulk free energy of each phase. 

In addition, we used our result for $K(V)$ to elucidate the role of the solute drag coefficient. Partial solute drag is predicted for a solidifying front in the context of the PFC model. As $V$ increases, the maximum solute drag decreases proportionately to the complete trapping velocity and solute relaxation time. The larger the solute relaxation parameter ($\gamma$), the lower the complete trapping velocity and therefore the smaller the amount of solute drag ($f$).  For fixed $\gamma$, the PFC model predicts a linear relationship between interface velocity and the total free energy for crystallization, consistent with recent MD simulations. It was found that the total available free energy for solidification and the maximum solute drag are velocity dependent, while the partial solute drag coefficient $f$ was independent of the velocity.  Model parameters such as those that alter the equilibrium segregation coefficient ($K_e$) were also examined. It was found that as $K_e$ increases, complete trapping occurrs at slower velocities due to lower driving forces required by the system. This also changes the maximum available solute drag, but, again, does not affect the solute drag coefficient $f$.

The results of this work comprise the first independent predictions of solute trapping and drag concepts emerging from a continuum theory that is fundamentally derived from a microscopic density functional theory. As a result, the analytical and numerical results presented here can be related to both thermodynamic material properties of the solid and liquid, as well as to the microscopic correlation properties of the melt from which crystallization occurs. 

The authors would like to acknowledge the National Science and Engineering Research Council of Canada (NSERC) for the finical support of this work.

\begin{thebibliography}{44}
\expandafter\ifx\csname natexlab\endcsname\relax\def\natexlab#1{#1}\fi
\expandafter\ifx\csname bibnamefont\endcsname\relax
  \def\bibnamefont#1{#1}\fi
\expandafter\ifx\csname bibfnamefont\endcsname\relax
  \def\bibfnamefont#1{#1}\fi
\expandafter\ifx\csname citenamefont\endcsname\relax
  \def\citenamefont#1{#1}\fi
\expandafter\ifx\csname url\endcsname\relax
  \def\url#1{\texttt{#1}}\fi
\expandafter\ifx\csname urlprefix\endcsname\relax\def\urlprefix{URL }\fi
\providecommand{\bibinfo}[2]{#2}
\providecommand{\eprint}[2][]{\url{#2}}

\bibitem[{\citenamefont{Gollub and Langer}(1999)}]{langer99}
\bibinfo{author}{\bibfnamefont{J.~P.} \bibnamefont{Gollub}} \bibnamefont{and}
  \bibinfo{author}{\bibfnamefont{J.~S.} \bibnamefont{Langer}},
  \bibinfo{journal}{Rev. Mod. Phys.} \textbf{\bibinfo{volume}{71}},
  \bibinfo{pages}{S396} (\bibinfo{year}{1999}).

\bibitem[{\citenamefont{Kittl et~al.}(2000)\citenamefont{Kittl, Sanders, Aziz,
  Brunco, and Thompson}}]{kittl00}
\bibinfo{author}{\bibfnamefont{J.~A.} \bibnamefont{Kittl}},
  \bibinfo{author}{\bibfnamefont{P.~G.} \bibnamefont{Sanders}},
  \bibinfo{author}{\bibfnamefont{M.~J.} \bibnamefont{Aziz}},
  \bibinfo{author}{\bibfnamefont{D.~P.} \bibnamefont{Brunco}},
  \bibnamefont{and} \bibinfo{author}{\bibfnamefont{M.~O.}
  \bibnamefont{Thompson}}, \bibinfo{journal}{Acta Mater.}
  \textbf{\bibinfo{volume}{48}}, \bibinfo{pages}{4797} (\bibinfo{year}{2000}).

\bibitem[{\citenamefont{Aziz and Kaplan}(1988)}]{aziz88}
\bibinfo{author}{\bibfnamefont{M.~J.} \bibnamefont{Aziz}} \bibnamefont{and}
  \bibinfo{author}{\bibfnamefont{T.}~\bibnamefont{Kaplan}},
  \bibinfo{journal}{Acta Metall.} \textbf{\bibinfo{volume}{36}},
  \bibinfo{pages}{2335} (\bibinfo{year}{1988}).

\bibitem[{\citenamefont{Galenko and Sobolev}(1997)}]{Galenko97b}
\bibinfo{author}{\bibfnamefont{P.}~\bibnamefont{Galenko}} \bibnamefont{and}
  \bibinfo{author}{\bibfnamefont{S.}~\bibnamefont{Sobolev}},
  \bibinfo{journal}{Phys. Rev. E} \textbf{\bibinfo{volume}{55}},
  \bibinfo{pages}{343} (\bibinfo{year}{1997}).

\bibitem[{\citenamefont{Ahmad et~al.}(1998)\citenamefont{Ahmad, Wheeler,
  Boettinger, and McFadden}}]{Ahmad98}
\bibinfo{author}{\bibfnamefont{N.}~\bibnamefont{Ahmad}},
  \bibinfo{author}{\bibfnamefont{A.}~\bibnamefont{Wheeler}},
  \bibinfo{author}{\bibfnamefont{W.}~\bibnamefont{Boettinger}},
  \bibnamefont{and} \bibinfo{author}{\bibfnamefont{G.}~\bibnamefont{McFadden}},
  \bibinfo{journal}{Phys Rev E} \textbf{\bibinfo{volume}{{ 58}}},
  \bibinfo{pages}{3436} (\bibinfo{year}{1998}).

\bibitem[{\citenamefont{K.~A.~Jackson and Gudgel}(2004)}]{jackson04}
\bibinfo{author}{\bibfnamefont{K.~M.~B.} \bibnamefont{K.~A.~Jackson}}
  \bibnamefont{and} \bibinfo{author}{\bibfnamefont{K.~A.}
  \bibnamefont{Gudgel}}, \bibinfo{journal}{J. Cryst. Growth}
  \textbf{\bibinfo{volume}{271}}, \bibinfo{pages}{481} (\bibinfo{year}{2004}).

\bibitem[{\citenamefont{Sobolev}(1995)}]{sob95}
\bibinfo{author}{\bibfnamefont{S.~L.} \bibnamefont{Sobolev}},
  \bibinfo{journal}{Phys. Rev. A} \textbf{\bibinfo{volume}{199}},
  \bibinfo{pages}{383} (\bibinfo{year}{1995}).

\bibitem[{\citenamefont{Sobolev}(1997)}]{sob97}
\bibinfo{author}{\bibfnamefont{S.~L.} \bibnamefont{Sobolev}},
  \bibinfo{journal}{Phys. Rev. E} \textbf{\bibinfo{volume}{55}}
  (\bibinfo{year}{1997}).

\bibitem[{\citenamefont{Cahn}(1962)}]{Cahn62}
\bibinfo{author}{\bibfnamefont{J.~W.} \bibnamefont{Cahn}},
  \bibinfo{journal}{Acta. Metall.} \textbf{\bibinfo{volume}{10}},
  \bibinfo{pages}{789} (\bibinfo{year}{1962}).

\bibitem[{\citenamefont{Hillert and Sundman}(1976)}]{Hil76}
\bibinfo{author}{\bibfnamefont{M.}~\bibnamefont{Hillert}} \bibnamefont{and}
  \bibinfo{author}{\bibfnamefont{B.}~\bibnamefont{Sundman}},
  \bibinfo{journal}{Acta Metall} \textbf{\bibinfo{volume}{24}},
  \bibinfo{pages}{731} (\bibinfo{year}{1976}).

\bibitem[{\citenamefont{Hillert and Schalin}(2000)}]{Hillert00}
\bibinfo{author}{\bibfnamefont{M.}~\bibnamefont{Hillert}} \bibnamefont{and}
  \bibinfo{author}{\bibfnamefont{M.}~\bibnamefont{Schalin}},
  \bibinfo{journal}{Acta Materialia} \textbf{\bibinfo{volume}{48}},
  \bibinfo{pages}{461} (\bibinfo{year}{2000}), \bibinfo{note}{qC 20100525}.

\bibitem[{\citenamefont{Smith and Aziz}(1994)}]{Smith94}
\bibinfo{author}{\bibfnamefont{P.}~\bibnamefont{Smith}} \bibnamefont{and}
  \bibinfo{author}{\bibfnamefont{M.~J.} \bibnamefont{Aziz}},
  \bibinfo{journal}{Acta Metall.} \textbf{\bibinfo{volume}{42}}
  (\bibinfo{year}{1994}).

\bibitem[{\citenamefont{Kittl et~al.}(1995)\citenamefont{Kittl, Aziz, Brunco,
  and Thompson}}]{kittl95}
\bibinfo{author}{\bibfnamefont{J.~A.} \bibnamefont{Kittl}},
  \bibinfo{author}{\bibfnamefont{M.~J.} \bibnamefont{Aziz}},
  \bibinfo{author}{\bibfnamefont{D.~P.} \bibnamefont{Brunco}},
  \bibnamefont{and} \bibinfo{author}{\bibfnamefont{M.~O.}
  \bibnamefont{Thompson}}, \bibinfo{journal}{J. Cryst. Growth}
  \textbf{\bibinfo{volume}{148}}, \bibinfo{pages}{172} (\bibinfo{year}{1995}).

\bibitem[{\citenamefont{Aziz and Boettinger}(1994{\natexlab{a}})}]{Aziz94}
\bibinfo{author}{\bibfnamefont{M.~J.} \bibnamefont{Aziz}} \bibnamefont{and}
  \bibinfo{author}{\bibfnamefont{W.~J.} \bibnamefont{Boettinger}},
  \bibinfo{journal}{Acta Metall. Mater.} \textbf{\bibinfo{volume}{{ 42 }}},
  \bibinfo{pages}{257} (\bibinfo{year}{1994}{\natexlab{a}}).

\bibitem[{\citenamefont{Yang et~al.}(2011)\citenamefont{Yang, Humadi, Buta,
  Laird, Sun, Hoyt, and Asta}}]{yangprl}
\bibinfo{author}{\bibfnamefont{Y.}~\bibnamefont{Yang}},
  \bibinfo{author}{\bibfnamefont{H.}~\bibnamefont{Humadi}},
  \bibinfo{author}{\bibfnamefont{D.}~\bibnamefont{Buta}},
  \bibinfo{author}{\bibfnamefont{B.~B.} \bibnamefont{Laird}},
  \bibinfo{author}{\bibfnamefont{D.}~\bibnamefont{Sun}},
  \bibinfo{author}{\bibfnamefont{J.~J.} \bibnamefont{Hoyt}}, \bibnamefont{and}
  \bibinfo{author}{\bibfnamefont{M.}~\bibnamefont{Asta}},
  \bibinfo{journal}{Phys. Rev. Lett.} \textbf{\bibinfo{volume}{107}},
  \bibinfo{pages}{025505} (\bibinfo{year}{2011}).

\bibitem[{\citenamefont{Agren}(1989)}]{Agren89}
\bibinfo{author}{\bibfnamefont{J.}~\bibnamefont{Agren}}, \bibinfo{journal}{Acta
  Metall.} \textbf{\bibinfo{volume}{37}}, \bibinfo{pages}{181}
  (\bibinfo{year}{1989}).

\bibitem[{\citenamefont{Humadi et~al.}(2013)\citenamefont{Humadi, Hoyt, and
  Provatas}}]{Humadi13}
\bibinfo{author}{\bibfnamefont{H.}~\bibnamefont{Humadi}},
  \bibinfo{author}{\bibfnamefont{J.~J.} \bibnamefont{Hoyt}}, \bibnamefont{and}
  \bibinfo{author}{\bibfnamefont{N.}~\bibnamefont{Provatas}},
  \bibinfo{journal}{Phys. Rev. E} \textbf{\bibinfo{volume}{87}},
  \bibinfo{pages}{022404} (\bibinfo{year}{2013}).

\bibitem[{\citenamefont{Caginalp}(1989)}]{Cag89}
\bibinfo{author}{\bibfnamefont{G.}~\bibnamefont{Caginalp}},
  \bibinfo{journal}{Phys.\ Rev.\ A} \textbf{\bibinfo{volume}{39}},
  \bibinfo{pages}{5887} (\bibinfo{year}{1989}).

\bibitem[{\citenamefont{Caginalp and Socolovsky}(1991)}]{Cag91}
\bibinfo{author}{\bibfnamefont{G.}~\bibnamefont{Caginalp}} \bibnamefont{and}
  \bibinfo{author}{\bibfnamefont{E.}~\bibnamefont{Socolovsky}},
  \bibinfo{journal}{SIAM J. Sci. Comp.} \textbf{\bibinfo{volume}{15}},
  \bibinfo{pages}{106} (\bibinfo{year}{1991}).

\bibitem[{\citenamefont{Karma and Rappel}(1998)}]{Karma98}
\bibinfo{author}{\bibfnamefont{A.}~\bibnamefont{Karma}} \bibnamefont{and}
  \bibinfo{author}{\bibfnamefont{W.~J.} \bibnamefont{Rappel}},
  \bibinfo{journal}{Phys. Rev. E} \textbf{\bibinfo{volume}{57}},
  \bibinfo{pages}{4323} (\bibinfo{year}{1998}).

\bibitem[{\citenamefont{Provatas et~al.}(1998)\citenamefont{Provatas,
  Goldenfeld, and Dantzig}}]{Provatas98}
\bibinfo{author}{\bibfnamefont{N.}~\bibnamefont{Provatas}},
  \bibinfo{author}{\bibfnamefont{N.}~\bibnamefont{Goldenfeld}},
  \bibnamefont{and} \bibinfo{author}{\bibfnamefont{J.}~\bibnamefont{Dantzig}},
  \bibinfo{journal}{Phys. Rev. Lett.} \textbf{\bibinfo{volume}{80}},
  \bibinfo{pages}{3308} (\bibinfo{year}{1998}).

\bibitem[{\citenamefont{Kim et~al.}(1999)\citenamefont{Kim, Kim, and
  Suzuki}}]{Kim99b}
\bibinfo{author}{\bibfnamefont{S.~G.} \bibnamefont{Kim}},
  \bibinfo{author}{\bibfnamefont{W.~T.} \bibnamefont{Kim}}, \bibnamefont{and}
  \bibinfo{author}{\bibfnamefont{T.}~\bibnamefont{Suzuki}},
  \bibinfo{journal}{Phys. Rev. E} \textbf{\bibinfo{volume}{{ 60}}},
  \bibinfo{pages}{7186} (\bibinfo{year}{1999}).

\bibitem[{\citenamefont{Karma}(2001)}]{Karma01}
\bibinfo{author}{\bibfnamefont{A.}~\bibnamefont{Karma}},
  \bibinfo{journal}{Phys. Rev. Lett} \textbf{\bibinfo{volume}{87}},
  \bibinfo{pages}{115701} (\bibinfo{year}{2001}).

\bibitem[{\citenamefont{Boettinger et~al.}(2002)\citenamefont{Boettinger,
  Warren, Beckermann, and Karma}}]{Boettinger02}
\bibinfo{author}{\bibfnamefont{W.}~\bibnamefont{Boettinger}},
  \bibinfo{author}{\bibfnamefont{J.}~\bibnamefont{Warren}},
  \bibinfo{author}{\bibfnamefont{C.}~\bibnamefont{Beckermann}},
  \bibnamefont{and} \bibinfo{author}{\bibfnamefont{A.}~\bibnamefont{Karma}},
  \bibinfo{journal}{Annu Rev. Mater. Res.} \textbf{\bibinfo{volume}{{ 32}}},
  \bibinfo{pages}{163} (\bibinfo{year}{2002}).

\bibitem[{\citenamefont{Echebarria et~al.}(2004)\citenamefont{Echebarria,
  Folch, Karma, and Plapp}}]{Echebarria04}
\bibinfo{author}{\bibfnamefont{B.}~\bibnamefont{Echebarria}},
  \bibinfo{author}{\bibfnamefont{R.}~\bibnamefont{Folch}},
  \bibinfo{author}{\bibfnamefont{A.}~\bibnamefont{Karma}}, \bibnamefont{and}
  \bibinfo{author}{\bibfnamefont{M.}~\bibnamefont{Plapp}},
  \bibinfo{journal}{Phys. Rev. E.} \textbf{\bibinfo{volume}{70}},
  \bibinfo{pages}{061604} (\bibinfo{year}{2004}).

\bibitem[{\citenamefont{Aziz and Boettinger}(1994{\natexlab{b}})}]{Azizbot93}
\bibinfo{author}{\bibfnamefont{M.~J.} \bibnamefont{Aziz}} \bibnamefont{and}
  \bibinfo{author}{\bibfnamefont{W.~J.} \bibnamefont{Boettinger}},
  \bibinfo{journal}{Acta Metall.} \textbf{\bibinfo{volume}{42}},
  \bibinfo{pages}{527} (\bibinfo{year}{1994}{\natexlab{b}}).

\bibitem[{\citenamefont{Provatas and Elder}(2010)}]{ProvatasElder10}
\bibinfo{author}{\bibfnamefont{N.}~\bibnamefont{Provatas}} \bibnamefont{and}
  \bibinfo{author}{\bibfnamefont{K.}~\bibnamefont{Elder}},
  \emph{\bibinfo{title}{Phase-Field Methods in Materials Science and
  Engineering}} (\bibinfo{publisher}{Wiley-VCH Verlag GmbH \& Co. KGaA},
  \bibinfo{year}{2010}), ISBN \bibinfo{isbn}{9783527631520}.

\bibitem[{\citenamefont{Elder et~al.}(2007)\citenamefont{Elder, Provatas,
  Berry, Stefanovic, and Grant}}]{Elder07}
\bibinfo{author}{\bibfnamefont{K.~R.} \bibnamefont{Elder}},
  \bibinfo{author}{\bibfnamefont{N.}~\bibnamefont{Provatas}},
  \bibinfo{author}{\bibfnamefont{J.}~\bibnamefont{Berry}},
  \bibinfo{author}{\bibfnamefont{P.}~\bibnamefont{Stefanovic}},
  \bibnamefont{and} \bibinfo{author}{\bibfnamefont{M.}~\bibnamefont{Grant}},
  \bibinfo{journal}{Phys. Rev. B.} \textbf{\bibinfo{volume}{{ 75}}},
  \bibinfo{pages}{064107} (\bibinfo{year}{2007}).

\bibitem[{\citenamefont{Elder et~al.}(2002)\citenamefont{Elder, Katakowski,
  Haataja, and Grant}}]{Elder02}
\bibinfo{author}{\bibfnamefont{K.~R.} \bibnamefont{Elder}},
  \bibinfo{author}{\bibfnamefont{M.}~\bibnamefont{Katakowski}},
  \bibinfo{author}{\bibfnamefont{M.}~\bibnamefont{Haataja}}, \bibnamefont{and}
  \bibinfo{author}{\bibfnamefont{M.}~\bibnamefont{Grant}},
  \bibinfo{journal}{Phys. Rev. Lett.} \textbf{\bibinfo{volume}{{ 88}}},
  \bibinfo{pages}{245701} (\bibinfo{year}{2002}).

\bibitem[{\citenamefont{Elder and Grant}(2004)}]{Elder04}
\bibinfo{author}{\bibfnamefont{K.~R.} \bibnamefont{Elder}} \bibnamefont{and}
  \bibinfo{author}{\bibfnamefont{M.}~\bibnamefont{Grant}},
  \bibinfo{journal}{Phys. Rev. E} \textbf{\bibinfo{volume}{70}},
  \bibinfo{pages}{051605} (\bibinfo{year}{2004}).

\bibitem[{\citenamefont{Berry et~al.}(2006)\citenamefont{Berry, Grant, and
  Elder}}]{Berry06}
\bibinfo{author}{\bibfnamefont{J.}~\bibnamefont{Berry}},
  \bibinfo{author}{\bibfnamefont{M.}~\bibnamefont{Grant}}, \bibnamefont{and}
  \bibinfo{author}{\bibfnamefont{K.~R.} \bibnamefont{Elder}},
  \bibinfo{journal}{Phys. Rev. E} \textbf{\bibinfo{volume}{{ 73}}},
  \bibinfo{pages}{031609} (\bibinfo{year}{2006}).

\bibitem[{\citenamefont{Stefanovic et~al.}(2006)\citenamefont{Stefanovic,
  Haataja, and Provatas}}]{Stefanovic06}
\bibinfo{author}{\bibfnamefont{P.}~\bibnamefont{Stefanovic}},
  \bibinfo{author}{\bibfnamefont{M.}~\bibnamefont{Haataja}}, \bibnamefont{and}
  \bibinfo{author}{\bibfnamefont{N.}~\bibnamefont{Provatas}},
  \bibinfo{journal}{Phys. Rev. Lett.} \textbf{\bibinfo{volume}{96}},
  \bibinfo{pages}{225504} (\bibinfo{year}{2006}).

\bibitem[{\citenamefont{Mellenthin et~al.}(2008)\citenamefont{Mellenthin,
  Karma, and Plapp}}]{Mellenthin08}
\bibinfo{author}{\bibfnamefont{J.}~\bibnamefont{Mellenthin}},
  \bibinfo{author}{\bibfnamefont{A.}~\bibnamefont{Karma}}, \bibnamefont{and}
  \bibinfo{author}{\bibfnamefont{M.}~\bibnamefont{Plapp}},
  \bibinfo{journal}{Phys. Rev. B} \textbf{\bibinfo{volume}{78}},
  \bibinfo{pages}{184110} (\bibinfo{year}{2008}).

\bibitem[{\citenamefont{Stefanovic et~al.}(2009)\citenamefont{Stefanovic,
  Haataja, and Provatas}}]{Stefanovic09}
\bibinfo{author}{\bibfnamefont{P.}~\bibnamefont{Stefanovic}},
  \bibinfo{author}{\bibfnamefont{M.}~\bibnamefont{Haataja}}, \bibnamefont{and}
  \bibinfo{author}{\bibfnamefont{N.}~\bibnamefont{Provatas}},
  \bibinfo{journal}{Phys. Rev. E} \textbf{\bibinfo{volume}{80}},
  \bibinfo{pages}{046107} (\bibinfo{year}{2009}).

\bibitem[{\citenamefont{Gr\'an\'asy et~al.}(2011)\citenamefont{Gr\'an\'asy,
  Tegze, T\'oth, and Pusztai}}]{Granasy10}
\bibinfo{author}{\bibfnamefont{L.}~\bibnamefont{Gr\'an\'asy}},
  \bibinfo{author}{\bibfnamefont{G.}~\bibnamefont{Tegze}},
  \bibinfo{author}{\bibfnamefont{G.~I.} \bibnamefont{T\'oth}},
  \bibnamefont{and} \bibinfo{author}{\bibfnamefont{T.}~\bibnamefont{Pusztai}},
  \bibinfo{journal}{Philosophical Magazine} \textbf{\bibinfo{volume}{91}},
  \bibinfo{pages}{123} (\bibinfo{year}{2011}).

\bibitem[{\citenamefont{Tegze et~al.}(2011)\citenamefont{Tegze, Toth, and
  Granasy}}]{granasyprl}
\bibinfo{author}{\bibfnamefont{G.}~\bibnamefont{Tegze}},
  \bibinfo{author}{\bibfnamefont{G.~I.} \bibnamefont{Toth}}, \bibnamefont{and}
  \bibinfo{author}{\bibfnamefont{L.}~\bibnamefont{Granasy}},
  \bibinfo{journal}{Phys. Rev. Lett.} \textbf{\bibinfo{volume}{106}}
  (\bibinfo{year}{2011}).

\bibitem[{\citenamefont{Athreya et~al.}(2006)\citenamefont{Athreya, Goldenfeld,
  and Dantzig}}]{Athreya06}
\bibinfo{author}{\bibfnamefont{B.~P.} \bibnamefont{Athreya}},
  \bibinfo{author}{\bibfnamefont{N.}~\bibnamefont{Goldenfeld}},
  \bibnamefont{and} \bibinfo{author}{\bibfnamefont{J.~A.}
  \bibnamefont{Dantzig}}, \bibinfo{journal}{Phys. Rev. E}
  \textbf{\bibinfo{volume}{74}}, \bibinfo{pages}{011601}
  (\bibinfo{year}{2006}).

\bibitem[{\citenamefont{Athreya et~al.}(2007)\citenamefont{Athreya, Goldenfeld,
  Dantzig, Greenwood, and Provatas}}]{Athreya07}
\bibinfo{author}{\bibfnamefont{B.~P.} \bibnamefont{Athreya}},
  \bibinfo{author}{\bibfnamefont{N.}~\bibnamefont{Goldenfeld}},
  \bibinfo{author}{\bibfnamefont{J.~A.} \bibnamefont{Dantzig}},
  \bibinfo{author}{\bibfnamefont{M.}~\bibnamefont{Greenwood}},
  \bibnamefont{and} \bibinfo{author}{\bibfnamefont{N.}~\bibnamefont{Provatas}},
  \bibinfo{journal}{Phys. Rev. E} \textbf{\bibinfo{volume}{76}},
  \bibinfo{pages}{056706} (\bibinfo{year}{2007}).

\bibitem[{\citenamefont{Huang et~al.}(2010)\citenamefont{Huang, Elder, and
  Provatas}}]{Huang10}
\bibinfo{author}{\bibfnamefont{Z.-F.} \bibnamefont{Huang}},
  \bibinfo{author}{\bibfnamefont{K.~R.} \bibnamefont{Elder}}, \bibnamefont{and}
  \bibinfo{author}{\bibfnamefont{N.}~\bibnamefont{Provatas}},
  \bibinfo{journal}{Phys. Rev. E} \textbf{\bibinfo{volume}{82}},
  \bibinfo{pages}{021605} (\bibinfo{year}{2010}).

\bibitem[{\citenamefont{Humadi}(2013)}]{Humadi_thesis}
\bibinfo{author}{\bibfnamefont{H.}~\bibnamefont{Humadi}}, Ph.D. thesis,
  \bibinfo{school}{McMaster Univeristy} (\bibinfo{year}{2013}).

\bibitem[{\citenamefont{Majaniemi and Grant}(2007)}]{Majaniemi07}
\bibinfo{author}{\bibfnamefont{S.}~\bibnamefont{Majaniemi}} \bibnamefont{and}
  \bibinfo{author}{\bibfnamefont{M.}~\bibnamefont{Grant}},
  \bibinfo{journal}{Phys. Rev. B} \textbf{\bibinfo{volume}{75}},
  \bibinfo{pages}{054301} (\bibinfo{year}{2007}).

\bibitem[{\citenamefont{P.~K.~Galenko and Elder}(2014)}]{private_galenko}
\bibinfo{author}{\bibfnamefont{F.~I.~S.} \bibnamefont{P.~K.~Galenko}}
  \bibnamefont{and} \bibinfo{author}{\bibfnamefont{K.~R.} \bibnamefont{Elder}},
  \bibinfo{journal}{Physica D} \textbf{\bibinfo{volume}{Preprint}}
  (\bibinfo{year}{2014}).

\bibitem[{\citenamefont{Baker and Cahn}(1971)}]{Cahn71}
\bibinfo{author}{\bibfnamefont{J.~C.} \bibnamefont{Baker}} \bibnamefont{and}
  \bibinfo{author}{\bibfnamefont{J.~W.} \bibnamefont{Cahn}}, in
  \emph{\bibinfo{booktitle}{Solidification}} (\bibinfo{year}{1971}),
  \bibinfo{number}{ASM}, p.~\bibinfo{pages}{23}.

\bibitem[{\citenamefont{Turnbull}(1962)}]{turnbull62}
\bibinfo{author}{\bibfnamefont{D.}~\bibnamefont{Turnbull}},
  \bibinfo{journal}{J. Phys. Chem.} \textbf{\bibinfo{volume}{66}}
  (\bibinfo{year}{1962}).

\end{thebibliography}

\end{document}